# Unsupervised learning segmentation for dynamic speckle activity images

**Lucía I. Passoni [a], Ana L. Dai Pra [b], Gustavo J. Meschino [a], Marcelo Guzman [a,c], Christian Weber [d,e], Héctor Rabal [d], Marcelo Trivi [d],**

[a] Laboratorio de Bioingeniería. Facultad de Ingeniería. Universidad Nacional de Mar del Plata. Juan B. Justo 4302. Mar del Plata. Argentina. isabel.passoni@gmail.com, gustavo.meschino@gmail.com
[b] Grupo de Inteligencia Artificial aplicado a Ingeniería. Facultad de Ingeniería. Universidad Nacional de Mar del Plata. Juan B. Justo 4302. Mar del Plata. Argentina. daipra@fi.mdp.edu.ar
[c] Laboratorio Laser. Facultad de Ingeniería. Universidad Nacional de Mar del Plata. Juan B. Justo 4302. Mar del Plata. Argentina. marcelo.guzman@fi.mdp.edu.ar
[d] Centro de Investigaciones Ópticas (CONICET La Plata-CIC). Facultad de Ingeniería. Universidad Nacional de La Plata. Calle 1 y 47 - La Plata. Argentina. marcelociop@yahoo.com.ar
[e] Facultad de Ciencias Agrarias y Forestales. Universidad Nacional de La Plata. Calle 60 y 119. La Plata. Argentina. cweber@ciop.unlp.edu.ar

**ABSTRACT**

This paper proposes the design of decision models based on Computational Intelligence techniques applied to image sequences of dynamic laser speckle. These models aim to identify image regions of biological specimens illuminated by a coherent beam coming from a laser. The field image is pseudo colored using a Self Organizing Map projection. This process is carried out using a set of descriptors applied to the intensity variations along time in every pixel of an image sequence. The models use descriptors selected to improve effectiveness, depending on the specific application. We present two examples of the application of the proposed techniques to assess biological tissues. The results obtained are encouraging and significantly improve those obtained using a single descriptor.

**Address all correspondence to:** Lucía I. Passoni, Laboratorio de Bioingeniería. Facultad de Ingeniería. Universidad Nacional de Mar del Plata. Juan B. Justo 4302. Mar del Plata. Argentina. isabel.passoni@gmail.com

1. **Introduction**

Dynamic speckle patterns have been used to assess issues of interest in different fields, like biology (seed analysis, animal sperm motility), medicine (capillary blood flow), industry (discovering bruising in fruits, paint drying, monitoring of ice cream melting, yeast bread, gels), and so on [1]. An extensive list of descriptors, computed with the intensity variation of the speckle pattern, has been presented in the literature [1-2]. Most of them achieve good performance to characterize the activity of the speckles in detecting, in some specific way, the dynamic characteristics of the sample phenomenon. However the efficiency is improved when using several descriptors instead of using a single one, as shown in [3] and [4]. Two approaches from the machine learning paradigm can be used to combine a set of descriptors: supervised and unsupervised learning systems.

Supervised learning seems to be an adequate approach when a particular known target is sought. The learning stage of the system is performed using labeled samples (belonging to already known groups). Once trained, the system is used to identify unknown samples; i.e. it is able to recognize unknown samples with similar behavior to those used for learning.

Recently, we proposed the use of a supervised learning procedure for dynamic speckle images based on the application of a Naïve Bayes statistical classifier comprising the use of several descriptors [3]. In that work, speckle patterns were analyzed by training a classifier with a set of descriptors labeled by an expert user. After the learning stage the system could assign a class (a label) to unknown samples belonging to the same domain of the training set.

In this new work, the goal is to identify similar dynamics among the samples without the use of previous knowledge. Unsupervised learning can be expected to distinguish between regions of different dynamics, it is, segmenting different speckle patterns according to the similarity of their associated dynamics. Which are the right descriptors for unpredicted unknown phenomena cannot be a priori determined.

In this approach, which can be considered a natural continuation of the former, an unsupervised Artificial Neural Network (ANN) uses known descriptors from the literature for its learning and classification steps. We propose a method to perform the visualization of sample regions using an image sequence of laser speckle patterns belonging to live specimens (not necessarily mobile), particularly fruits and corn seeds. These sequences are used as input to a Self-Organizing Map (SOM) [5]. This type of neural network has also been used to characterize a chemotaxis assay; regions were neatly differentiated according to the bacterial motility within the sample [4]. In [6], SOMs were proposed as clustering methods in those cases where the sensitivity of the activity measurement of dynamic speckle images needed to be improved.

A multi-descriptor approach driven by the dynamics of the phenomenon itself does not require the assistance of an expert in the learning stage and has a better performance than other approaches that propose single descriptor analysis [6-7]. The contribution of this work is mainly the proposal of a technique for the coloring of regions of interest in the domain of the biological specimens' samples, as the quality assessment of corn seeds and the discovery of bruising in apples.

In this way, it is possible that this approach could discover by itself new features in the samples independently of the experts' previous knowledge.

## 2. Materials and Methods

*2.1 Equipment setup, signal acquirement, and feature extraction*

Figure 1 shows the experimental setup for acquisition and storage of the so called subjective images of a sample showing dynamic speckle. An expanded He-Ne laser (632.8nm and 30mW) illuminates the sample under study. A CCD camera connected to a frame grabber registers a sequence of intensity images (8 bits) and stores them into the personal computer.

The CCD objective was focused on the surface of the sample and the laser intensity is carefully kept constant during data acquisition. Speckle size was adjusted so that they were well resolved by the CCD pixels.

To evaluate the dynamics within stationary periods, images sequences were registered using a 4 Hz sampling frequency, as shown in Figure 2. The intensity from each image pixel was converted into a time series to be processed by computing different descriptors. So, the feature extraction was performed over the time series of intensity level in a pixelwise basis, computing numerical descriptors for every pixel location.

There are many descriptors that have been developed to characterize biospeckle [1]. We have experimented successfully the discovering of region of interests in fruits and seed with a set of descriptors in the time domain and time-frequency, we select those that provide the best discrimination of the areas of interest, such as: the Average of Subtraction of Consecutive Images, the Fujii descriptor, the Dynamic Range Descriptor, the Fuzzy Granular Descriptor and the Shannon Entropy of the Discrete Wavelet Transform to feed the neural network.

*2.2 Feature extraction: descriptors*

The descriptors equations are showed, considering $I_k(x,y)$ as the intensity level in the image pixel location *(x,y)* of the image *k*, with *k=1...N*

*2.2.1 Subtraction Average of consecutive pixel intensities*

One of the simplest descriptor is the Subtraction Average (SA) of two consecutive elements of the time speckle pattern [1].

$$SA = \sum_{k=1}^{N-1} \left| I_k(x,y) - I_{k+1}(x,y) \right| \Big/ N-1 \tag{1}$$

*2.2.2 Fujii Descriptor*

$$F(x,y) = \sum_{k=1}^{N} \frac{I_k(x,y) - I_{k-1}(x,y)}{I_k(x,y) + I_{k-1}(x,y)} \tag{2}$$

This measure has proved to be a very good estimation of activity when there are no spatial variations on the illumination of the sample [8].

*2.2.3 Dynamic Range Descriptor*

Dynamic Range descriptor was computed as the difference between the maximum and the minimum value of the intensity in each evaluated time series. The potential of this feature lies in its speed and ability to discriminate regions of coarse different activity [2].

$$DR = \max_{k=1,N}\{I(x,y)_k\} - \min_{k=1,N}\{I(x,y)_k\} \tag{3}$$

*2.2.4 Fuzzy Granular Descriptor*

The Fuzzy Granular algorithm is based on granular computing. It can be applied to both stationary and non-stationary cases, allowing monitoring the phenomenon in almost real time. According to the histogram of the image stack, different types of granules are identified; they are detected and counted, giving a descriptor that weights the series changes through the number of granules in a fixed time lapse [9].

The fuzzy sets theory, making reference to vague and overlapped concepts, allows defining granules with this property. To generate information granules several fuzzy sets are defined into the intensity values domain of the time speckle series. For intensity values $I_k(x,y)$, a fuzzy set is defined by a membership function $\mu_c(I_k(x,y))$ that takes gradual values in the real interval [0,1] (Eq. 4).

Trapezoidal functions $\mu_{dark}$, $\mu_{medium}$ and $\mu_{light}$ with media overlapping are adopted, where:

$$\mu_c(I(x,y)) \in [0,1], \text{ with } c \in \{dark, medium, light\} \quad (4)$$

Each granule is defined as a continuous sequence of elements belonging to the same intensity concept. The Fuzzy Granular Descriptor is the result of applying Eq. 5.

$$Q_N = \left( \sum_{c=1}^{3} suc_{k,c} \left[ \mu_c \left( I_k(x,y) \right) \right] \right) / N, \quad k = 1,2...,N$$

$$suc_{k,c} = \begin{cases} 1 & \text{if } \mu_c \left( I_{k-1}(x,y) \right) \neq 0 \wedge \mu_c \left( I_k(x,y) \right) = 0 \\ 0 & \text{else} \end{cases} \quad (5)$$

*2.2.5 Wavelet Entropy based Descriptor*

According to information theory, entropy is a relevant measure of order and disorder in a dynamical system. By using entropy, no specific distribution needs to be assumed. The spectral entropy as defined from the Fourier power spectrum shows a natural approach to quantify the degree of order of a complex signal, indicating the spread level of the signal power spectrum. The stationary condition to apply the Fourier transform (FT) is not ensured in the time speckle patterns. To deal with these limitations, time evolving entropy can be defined from a time-frequency representation of the signal as provided by the discrete wavelet transform (WT). Previous works have reported encouraging results for the identification of biological dynamics with this tool [10-12]. In order to study the biospeckle, the time series speckle patterns were divided in NT temporal windows of length L. The energy of the detail j of WT decomposition, using Daubechies wavelet (order=2), was applied to obtain the window Shannon entropy ($S_{WT}$). The value was assigned to the window central point.

$$S_{WT}^{(i)} = -\sum_{j<0} \frac{E_j^{(i)}}{\sum_j E_j^{(i)}} \cdot \ln\left[\frac{E_j^{(i)}}{\sum_j E_j^{(i)}}\right] \qquad (6)$$

The mean energy of the j coefficients in each window i ($C_{k,j,i}$), is obtained using eq.(6):

$$E_j^{(i)} = \frac{1}{N_j} \sum_{k=0}^{(L/2^j)-1} \left|C_{k,j,i}\right|^2 \qquad (7)$$

with $i = 1, \ldots, N_T$, and $C_{k,j,i}$ can be interpreted as the local residual errors between successive signal approximations.

The $S_{WT}$ entropy has been proposed in previous works to characterize the biospeckle phenomenon in images sequences that show inhomogeneous activity within different regions. A new image is generated with $S_{WT}$ values.[10]

*2.3 Pseudo-coloring by Self Organized Maps*

The Self Organizing Map (SOM) proposed by Kohonen is a popular non supervised neural network model [5]. The SOM quantizes the data space of training data and simultaneously it performs a topology-preserving projection of the data space onto a regular neuron (or cell) grid. SOM structure is usually a regular 2-dimensional grid of neurons, though they can be arranged in 1-dimensional (line) or 3-dimensional (space). Considering $D$-dimensional input data, each neuron $i$, is connected to the inputs by $D$ weights. From another point of view, these weights can be seen as cells built up with vectors of $D$ dimensions. This set of reference vectors is called the SOM *codebook*. The map cells are related each other by a neighborhood function. There are no weights that explicitly interconnect the neurons.

During each training step, one sample vector from the input data set is taken and a similarity measure is computed between the input vector and all the codebook vectors. The cell whose weight vector has the greatest similarity with the input sample is selected as the *Best-Matching Unit* (BMU). The similarity is defined by means of a distance measure, typically Euclidean distance.

After finding the BMU, the *codebook* is updated. The reference vectors of the BMU and its topological neighbors (according to the neighborhood function) are changed in order to be "closer" to the input vector in the input space. This adaptation procedure stretches the BMU and its topological neighbors towards the sample vector. The adaptation is given by:

$$W_j(n+1) \leftarrow W_j(n) + \eta(n) h_{ji}(n) \left[ X(n) - W_j(n) \right], \tag{8}$$

where $n$ is the iteration number, $j$ is the neuron index that is considered in the current iteration, $W_j$ is the prototype vector of cell $j$, $\eta(n)$ is a learning rate, $h_{ji}(n)$ is the neighborhood function defined centered on BMU, and $X(n)$ is the input data presented. Usually, both learning rate and the neighborhood function radius are decreasing as iterations progress.

Once trained, the SOM offers different ways to be visualized and analyzed. A matrix of distances between the codebook vectors of the cells and their neighbors is widely used [13-15]. Data samples can be projected onto the SOM by their BMU. Similar data will be projected in near cells.

In order to evaluate the quality of the map, two kinds of errors are considered: the quantization error and topographic error [12, 13]. They tend to minimize when the map vectors perform an organized projection of the training pattern according to a similarity criterion. Quantization error is computed as:

$$E_Q = \frac{1}{N} \sum_{i=1}^{N} \|x_i - m_i\|, \tag{9}$$

where $x_i, i = 1, 2..., N$ are the training data, $m_i$ is the BMU corresponding to datum $x_i$, and $N$ is the number of data.

Topographic error is helpful to assess whether the data topology was preserved after training, and it is computed as:

$$E_T = \frac{1}{N} \sum_{i=1}^{N} u(x_i), \tag{10}$$

where $u(x_i) = 1$ if the BMU for datum $x_i$ is not equal to the second BMU y $u(x_i) = 0$ if it is equal.

Once the map is properly trained, colors can be heuristically assigned to cells according to the distance between prototype vectors. The color coding is such that topological nearby cells will have similar colors and those far according to this criterion will have distinct colors. Using this colored map, a color can be assigned to each input data according their BMUs. [16]

To assign colors, we followed the next steps:

- Codebook matrix is projected into a 2-dimensional space by Principal Components Analysis (PCA) analysis.
- PCA codebook coordinates are scaled in the [0, 1] range.
- Colors are assigned to each cell of the PCA codebook coordinates using a RGB palette, selected to give a well differenced picture at most of LCD monitors.

**3. Experiments**

*3.1 Quality corn seed assessment*

Over the years, corn was finding different uses depending on the physicochemical composition that defines the type of grain. The quality of corn grain is associated with both, physical composition, which determines the texture and hardness, and chemical composition, which defines the nutritional and technological properties. The corn grain consists of four main parts, where the endosperm is 80-85%, 10-12% embryo, the pericarp 5-6% and 2-3 percent aleurone. The endosperm chemical composition is what sets different grain shapes and physical characteristics, which enable the commercial rates [17].

There has been interest among seed corn buyers about the differ-ences in the type of starch found in hybrids. What these discussions are referring to is the amount of floury (also called soft or dent) endosperm versus vitreous (also called hard or flint) endosperm [18].

Currently the method is the flotation test according to the standard of the Secretary of Agriculture, Livestock, Fisheries and Food in Argentinean Republic. An aqueous solution of sodium nitrate is used, achieving a specific gravity of 1.25 to water kept at a temperature of 35 °C.

This method allows comparing the density of various batches of corn kernels; it is based on the principle that the hard grains are of greater density and therefore such grains float in lower proportion than the grains of lower density in the solution of sodium nitrate.

Quantification of floating grains cannot determine the amount of floury and starch endosperm present in a sample. That is why we propose to deal with optical technologies to determine the possibility of using the method of speckle in such disquisition. The aim of the proposed process is to provide a tool to be used jointly with digital image processing methodologies to determine areas of the corresponding endosperm fractions. The activity of the endosperm is focused on its two majority parties (floury and vitreous endosperm), and the issue of successfully automating the identification of these areas would be of potential importance for trade and industrialization.

Previous work has been done to evaluate the corn seed viability using dynamic speckle patterns jointly with a descriptor that identifies the region of the live embryo [10,19]

*3.2 Bruising detection in Apples*

Impact damage is one of the quality defects that are most likely to be found in apples and pears. It produces significant quality losses. The impacts and pressures that don't break the epidermis but cause damage into the pulp of the fruit are called mechanical damage. It produces progressive color changes, being perceptible after a few days after a hit. Color changes are due to physical changes in the texture of the tissue and sometimes due to chemical alteration of the impacted zone, and they depend on the variety and the tissue structure of each fruit: skin thickness, pigmentation, etc.

A known apple variety, *Red Delicious*, is a sensitive one. Its external color is such that it prevents the mechanical impact damage hits to be clearly identifiable.

While this problem has been addressed using a single descriptor at a time [20-22]., this paper proposes the use of a set of descriptors and self-organizing map for improving automatic recognition of regions of interests like bruises.

The experiment was carried out hitting apples without any damage, *Red delicious* type, with a steel sphere (diameter = 21.9 mm and weight = 133.6 g) from 20 cm height. The bruising produced could not be detected by a visual inspection.

Several images short sequences (less than 100 images) as those shown in fig. 2 were registered: immediately after the hit in the same experimental conditions. Three already known descriptors were selected to train the SOM: Dynamic Range, Fujii and Wavelet Entropy.

*3.3 Proposed processing pipeline*

Summarizing, the sequence of steps of the method we propose is:

**STEP 1:**

Given the setup, acquire a sequence of images. Each pixel will be a temporal series.

**STEP 2:**

Compute features presented in Section 2.2 for each pixel of the image stack, arranging them in vectors.

We obtained a dataset with one vector for each pixel.

**STEP 3:**

Train a SOM with the dataset obtained in the previous step.

SOM dimensions (the number of cells and their arrangement) were chosen from a growing algorithm that weights the quantification and topographic errors. Training was stopped when these quality parameters achieved stationary behavior.

The SOM was initialized as a 2D hexagonal grid with Gaussian neighborhood. Training patterns were normalized to the [0, 1] range and the initialization was done using a lineal mapping of the training data. The SOM toolkit for Matlab®, from the Laboratory of Computer and Information Science (CIS) at the Helsinki University of Technology, was used.

**STEP 4:**

Visualize the trained SOM choosing a proper coloring code for the codebook as explained in section 2.3, so pixels with similar temporal behaviors will be colored with similar colors and conversely, pixels with different temporal behaviors will be colored with different colors. Identify the areas of the map according activity of the temporal series.

**STEP 5:**

Create a new color image, assigning the color of each pixel taking the color of the SOM cell who is the BMU for the vector of the pixel.

The processing pipeline is represented on Figure 3.

## 4. Results

### *4.1 Quality corn seed assessment*

Temporal descriptors were computed for all samples. Thus the Subtraction Average of time consecutive pixel intensities, the Dynamic Range and the Fuzzy Granular descriptors were obtained for the stack of 300 images of 400x400 pixels, achieving 1,600,000 patterns vectors. In order to balance the type of information, part of the image background was trimmed, given that in the acquisition stage the background was rather oversized. Finally images of 400x300 pixels were computed. Figure 4 shows images corresponding to descriptors of a specimen. Note that the discrimination ability in the four identified areas (background, embryo, floury endosperm and vitreous endosperm) is not achieved with any of them.

In order to determine the SOM dimension, a growing configuration for increasing dimensions of the grid size was proposed, with a stopping criterion of minimizing the topographic error, obtaining a good projection of similar vectors in neighboring cells. Linear codebook initialization was performed. The dimension was determined as a 10x10 cell array. The three inputs vectors components were the three above mentioned descriptor, all of them normalized according with their variance.

Definition of map size was made disposing of all data generated by the 10 trials for a total of 90,000 vectors of three variables, generated by the images of 300x300 pixels. We carried out a scheme of training and test by a cross-validation process. Since there were 10 different experimental tests, we performed a leave-one-out scheme, with 10 runs, training with 9 cases and testing with the left out case. The errors obtained after training the SOM were 0.2499 for the quantification error and 0.03960 for the topographic error.

The 10x10 codebook vectors are colored according to the distance between prototype vectors with a RGB palette as shown in Figure 5 (a). In Figure 5 (b) the values of the two principal components of the Principal Component Analysis (PCA) of the SOM codebook are shown. Figure 5(b) shows a projection of the map codebook using the two main components, where the circles are the component values, and the lattice segments are proportional to the distances between them. Note that the projections of the four vertices of the colored map of Figure 5(a) are distant in Figure 5(b). So the projections of corners of Fig 5(a): the blue upper right, the purple lower right, the light green upper left and the orange lower left are also well separated in Fig 5(b).

Figure 6 (a) and (b) show two corn seed images colored according with the labeled regions. To evaluate the goodness of the proposal we used expert's opinion, who considered, observing the prepared seeds and the pseudo-colored images of the processed speckle patterns that endosperm areas were properly identified. Also he noted that the methodology could be considered appropriated to automate the calculation of the fractions of floury and vitreous endosperm of seeds.

As shown in Figure 6, pseudo coloring of corn flinty seeds enables differentiating the background region (blue color), that is highly spaced in the codebook from live parts. Endosperm starchy region is distinguished perfectly in a purple coloration, as also the region of the embryo (yellow-green zone) and the vitreous endosperm (orange zone). Thus, the four classes can be perfectly discriminated, a fact that was not possible using only one of the time domain descriptors.

*4.2 Brusing in Apples*

In order to train the SOM, a set of image sequences were chosen. In these images, an object with inert surface was added (a piece of metal), where there is no formation of specks, in addition to regions of the bruised area. Thus the neural network was trained with three types of areas: healthy apple tissue, bruised tissue and inert zone, using feature vectors with three components: Dynamic Range, Fujii and Wavelet Entropy resultant values. Then, it was consulted with the results of other experiments: non- bruised apples and bruised apples, without the inert region.

The growing algorithm generated a 25x25 cells map, and the errors obtained after training the SOM were 0.035 for the quantification error and 0.076 for the topographic error. The unified distance matrix (U-matrix) was obtained (Figure 7), where three clusters can be neatly discovered (inert region, bruised area and healthy tissue) with low distance values (blue tones). By using previous knowledge of the experiment, we identified the clusters as belonging to the different regions. We conclude that the chosen descriptors are adequate to "discover" the existence of the bruising.

We also obtained another visualization of the internal state of the SOM in order to find out new conclusions. A new colored map is obtained, where cells are colored according to their similarity (Cell prototype vectors Colored by Similarity- CCS), that can be observed in Figure 8 (a). This representation confirms the conclusions obtained from the U-matrix: the "bruised region" upper right corner (yellow hues) in agreement with the bruised region in the U-matrix. The blue spot (bottom right) goes for the inert region, while the brown-green area points the healthy tissue.

By taking the BMU for each pixel a pseudo-colored image was obtained. When computing BMU's cells from the training sample, an image is mapped with the CCS color where each BMU impacts. A pseudo-colored image is obtained, which is shown in Figure 8 (b).

The trained SOM was recalled with a test set of speckle sequences that were measured under the same protocol. Figure 9 shows the result images of the test set, where the existence of yellow regions discovers bruising in different samples, if they exist. In (a) and (b) a case with enhanced bruise region is presented and in (c) a case with no bruise region is shown. The latter case is the result of a speckle sequence acquired without bruising.

**5. Conclusions**

In this work we proposed the use of Self-Organizing Maps to visually identify diverse regions in biological specimens like bruising in apples and endosperms types in corn seeds. The methodology proposed is based in an optic approach together with a multi descriptor approach to characterize dynamic speckle images. The approach has been shown to learn itself from the descriptors values to distinguish regions with similar dynamics in unsupervised way.

The corn seed case addresses the segmentation of regions of the endosperm embryo with the aim to design an automated process to compute fractions of interest. In this case the involved descriptors are computed within the time domain generating a lower computational cost compared with the computation of descriptors in the frequency domain. It is worth noting that the biggest cost in a real time process is due to the computation of descriptors.

The results obtained by processing a set of specimens of flint corn are very encouraging. This proposal is novel in the field of agricultural technology and will aim to provide a methodology for assessing the quality of corn, with industrial interest, based on the content of the endosperm.

In the case of the bruised apples, we proposed a methodology to detect the bruised region when it is not yet visible. It has been tested the robustness and reliability of the method in so far as having the SOM trained with a video sequence, has allowed the discovery of bruises in samples from other experiments. This process improves the results obtained with single descriptors, as in most of the previously published works. This is particularly true when the registered sequences are not very long, it is, less than 100 images.

It must be considered that as in practice the model must be previously trained, according to the experiment features, so its recall demands a low computational cost. This characteristic allows its inclusion into an on-line process for recognition of regions of interest into an artificial vision system.

The procedure used in these examples can be straightforwardly extended for testing other active samples, each with its characteristic dynamics requiring the choice of the adequate descriptors. The presence of new phenomena eventually could be discovered in this way.

As future work, we plan to assess the results using the same pipeline in other biological time-changing phenomena.

**NOTE:** Preliminary results related to this work were published unrefereed in "Dynamic laser speckle: decision models with computational intelligence techniques", Marcelo Guzman, Gustavo J. Meschino, Ana L. Dai Pra, Marcelo Trivi  Lucía I. Passoni, Héctor Rabal, Speckle 2010: Optical Metrology, edited by Armando Albertazzi Goncalves Jr., Guillermo H. Kaufmann, Proc. of SPIE Vol. 7387, 738717. (2010).

**Acknowledgments**

This work was supported by CCT La Plata Consejo Nacional de Investigaciones Científicas y Técnicas (CONICET), Comisión de Investigaciones Científicas de la Provincia de Buenos Aires, by Facultad de Ingeniería, La Plata National University, by Facultad de Ingeniería, Mar del Plata National University and by a Grant PICT 2008-1430, Agencia Nacional de Promoción Científica y Tecnológica, Argentina.

# References

1. H. J. Rabal, R. A. Braga (Eds.), *Dynamic Laser Speckle and Applications*, CRC Press, (2008).

2. G. H. Sendra, A. L. Dai Pra, L. I. Passoni, R. Arizaga, H. J. Rabal, M. Trivi, "Biospeckle descriptors: a performance comparison", *Proc. SPIE* 7387, 73871K (2010), DOI:10.1117/12.870682

3. L. Passoni, H. Rabal, G. Meschino, M. Trivi. "Probability mapping images in dynamic speckle classification". *Applied Optics* 52 (4), 726-733, (2013).

4. G. Meschino, S. Murialdo, L. I. Passoni, H. J. Rabal, M. Trivi; "Biospeckle image stack process based on artificial neural networks", Engineering in Medicine and Biology Society (EMBC), 2010 Annual International Conference of the IEEE, pp.4056-4059, Aug. 31 2010-Sept. 4 2010 doi: 10.1109/ IEMBS.2010.5627620

5. Kohonen, T.: *Self-Organizing Map*. Springer-Verlag. (1995).

6. P. Etchepareborda, A. Federico, G. Kaufmann, "Sensitivity evaluation of dynamic speckle activity measurements using clustering methods," *Applied Optics*, 49, 3753-3761 (2010).

7. P. Engelbrecht Andries, *Computational Intelligence: An Introduction*. Chichester: John Wiley (2007).

8. H. Fujii, T. Asakura, K. Nohira, Y. Shintomi, y T. Ohura, "Blood flow observed by time varying laser speckle", *Optics Letters* 10 (3), 104-106 (1985).

9. A. L. Dai Pra, L. I. Passoni, H. Rabal, "Evaluation of laser dynamic speckle signals applying granular computing", *Signal Processing*, 89, 266-274, (2009). {doi:10.1016/j.sigpro.2008.08.012.

10. L. Passoni, A. L. Dai Pra, H. J. Rabal, M. Trivi, R. Arizaga, "Dynamic speckle processing using wavelets based entropy" *Optics Communications*, 246, 219-228 (2005).

11. R.A. Braga Jr., G.W. Horgan, A.M. Enes, D. Miron, G.F. Rabelo, J.B. Barreto Filho, "Biological feature isolation by wavelets in biospeckle laser images", *Computers and Electronics in Agriculture*, 58, 123-132, (2007).

12. C.M.B. Nobre, R.A. Braga Jr., A.G. Costa, R.R. Cardoso, W.S. da Silva, T. Sáfadi, "Biospeckle laser spectral analysis under Inertia Moment, Entropy and Cross-Spectrum methods", *Optics Communications*, 282, 2236-2242, (2009).

13. Juha Vesanto, "SOM-based data visualization methods", *Intelligent Data Analysis*, 3, 111-126, (1999).

14. Juha Vesanto and Mika Sulkava "Distance matrix based clustering of the SelfOrganizing Map". In Dorronsoro, J. R., editor, Proceedings of the 12th International Conference on Artificial Neural Networks (ICANN 2002). Madrid, Spain, 2730 August 2002. *Lecture Notes in Computer Science*, volume 2415, pages 951956.

15. Vesanto, J., Sulkava, M., and Hollmen, J. On the Decomposition of the Self-Organizing Map Distortion Measure. *Proceedings of the Workshop on Self-Organizing Maps* (WSOM'03), pp 11-16, Hibikino, Kitakyushu, Japan, September 2003

16. M. C. de Matos, K. J. Marfurt, P. R. S. Johann, "Seismic interpretation of self-organizing maps using 2D color displays". *Rev. Bras. Geof.* [online], 28, 631-642. (2010).

17. S. M. Drury, T. L. Reynolds, W. P. Ridley, N. Bogdanova, S. Riordan, M. A. Nemeth, R. Sorbet, W. A. Trujillo, M. L. Breeze, "Composition of Forage and Grain from Second-Generation Insect-Protected Corn MON 89034 Is Equivalent to That of Conventional Corn (Zea mays L.)". *J. Agric. Food Chem.*, 56, 4623–46302. (2008).

18. Bill Mahanna and Ev Thomas https://www.pioneer.com/home/site/us/menuitem.b8381b50868d5c8176f576f5d10093a0/ April 2012

19. R. A. Braga, I. M. D. Fabbro, F. M. Borem, G. Rabelo, R. Arizaga, H. J. Rabal, M. Trivi, "Assessment of seed viability by laser speckle techniques", *Biosystems Engineering* 86 (3) 287-294, (2003) { 294. doi:10.1016/j.biosystemseng.2003.08.005.

20. M. Pajuelo, G. Baldwin, H. Rabal, N. Cap, R. Arizaga, M. Trivi, "Biospeckle assessment of bruising in fruits", *Optics and Lasers in Engineering*, 40, 13-24, (2003) { 24, <ce:title>Optics in Latin America part II</ce:title>. doi:10.1016/S0143-8166(02)00063-5.

21. G.G. Romero, C.C. Martinez, E.E. Alanís, G.A. Salazar, V.G. Broglia, L. Álvarez, "Bio-speckle activity applied to the assessment of tomato fruit ripening", *Biosystems Engineering*, 103, 116-119, (2009). 10.1016/j.biosystemseng.2009.02.001.

22. A. Zdunek, L. Muravsky, L. Frankevych, K. Konstankiewicz, "New nondestructive method based on spatial-temporal speckle correlation technique for evaluation of apples quality during shelf-life". *Int. Agrophy*., 21, 305-310. (2007).

**Figure Captions**

**Figure 1:** Optical setup.

**Figure 2:** Laser speckle sequence of images.

**Figure 3:** Processing pipeline for the pseudo-coloring using SOM.

**Figure 4:** Descriptor Images and the four region of interest: Vitreous Endosperm (VE), Floury Endosperm (FE), Embryo region (ER) and background (BG). Left: subtraction Average Descriptor; middle: dynamic Range Descriptor; right: Fuzzy Granular Descriptor

**Figure 5:** (a): SOM colored map. (b) SOM Colored Principal Components.

**Figure 6:** (a) and (b) two specimens of flint corn seed, pseudo-colored according to the regions of interests given by the SOM codebooks.

**Figure 7:** Unified Distance Matrix (U-matrix) 3D Representation of the distance matrix.

**Figure 8:** (a) Cell prototype vectors.Colored by Similarity- CCS Map colored by similarity of (b) Image of the training sample clearly showing the bruised region in yellow color.

**Figure 9:** SOM evaluations of test samples. (a) ,(b) two different bruised apples, c) Non-bruised apple. Bruised area is colored as bright yellow.

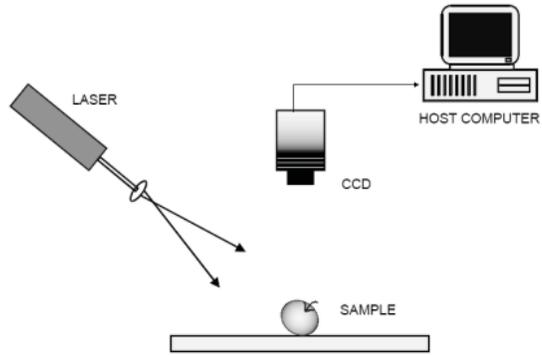

Figure 2. Optical setup

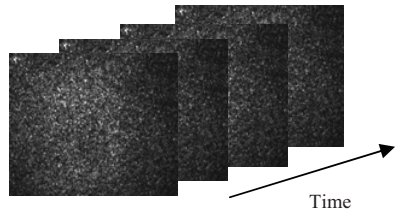

Figure. 2. Laser speckle sequence of images.

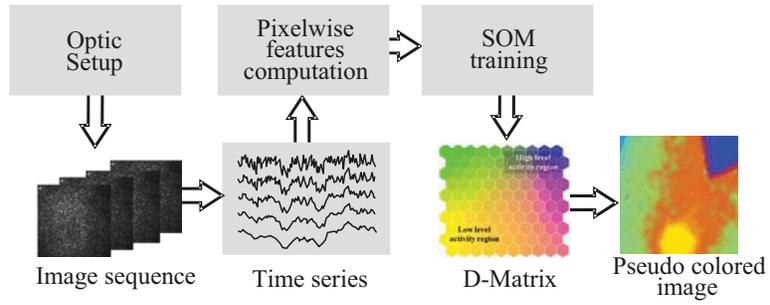

Figure 3. Processing pipeline for the pseudo-coloring using SOM.

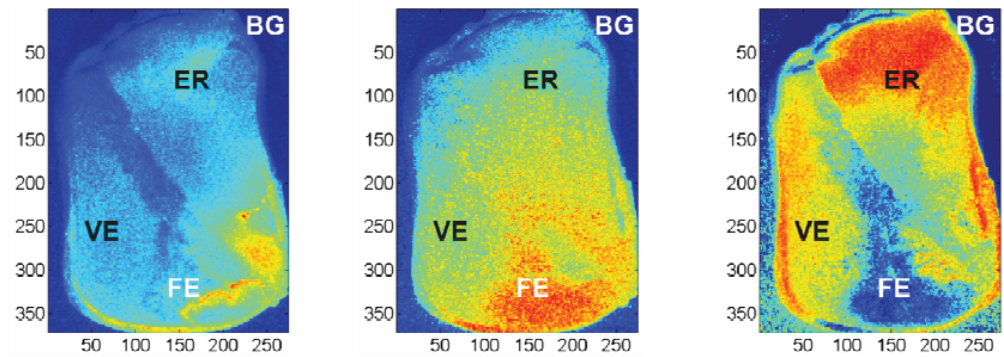

**Figure. 4.** Descriptor Images and the four region of interest: Vitreous Endosperm (VE), Floury Endosperm (FE), Embryo region (ER) and background (BG). Left: subtraction Average Descriptor; middle: dynamic Range Descriptor; right: Fuzzy Granular Descriptor.

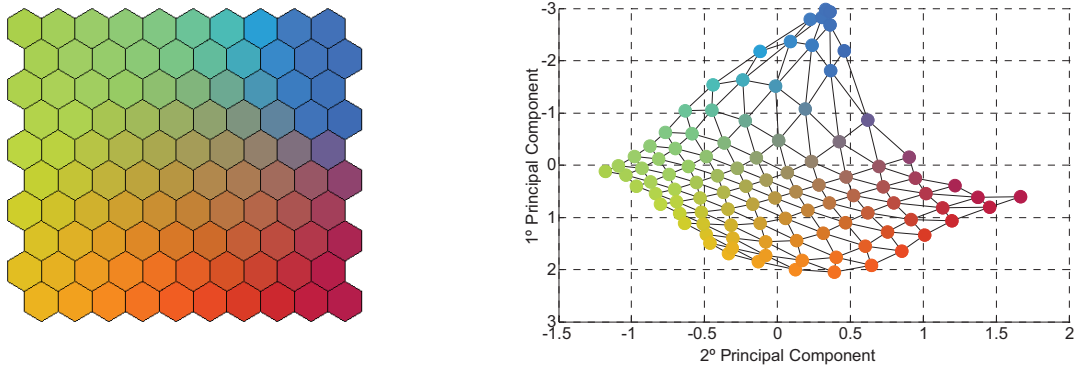

Figure 5 (a): SOM colored map. (b) SOM Colored Principal Components.

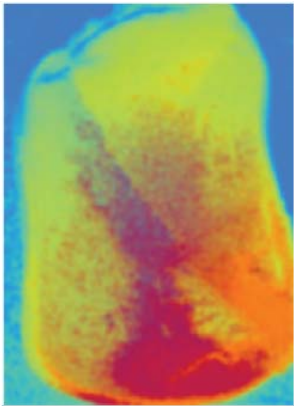

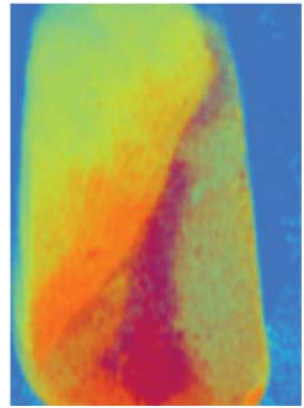

(a)                                                                                        (b)

**Figure 6.** (a) and (b) two specimens of flint corn seed, pseudo-colored according to the regions of interests given by the SOM codebooks.

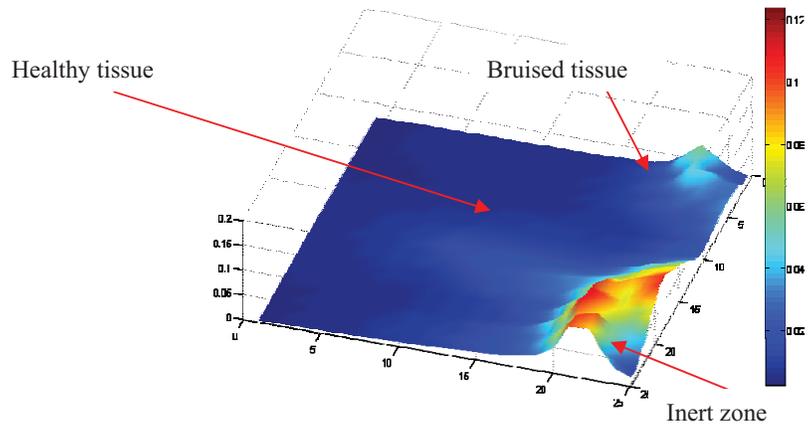

**Figure 7**.Unified Distance Matrix (U-matrix) 3D Representation of the distance matrix

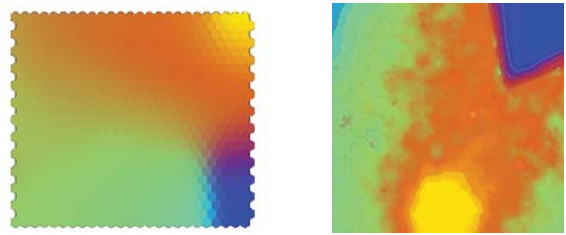

(a)　　　　　　　　　　　　　(b)
**Figure 8**: (a) Cell prototype vectors.Colored by Similarity- CCS Map colored by similarity of (b) Image of the training sample clearly showing the bruised region in yellow color.

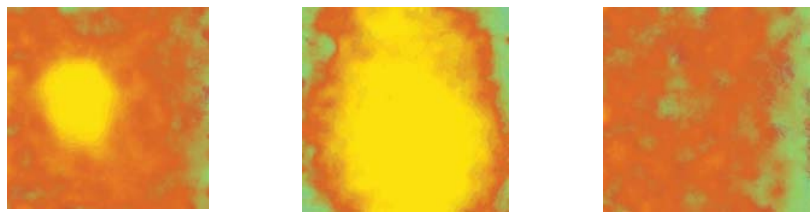

(a)　　　　　　　　(b)　　　　　　　　(c)

**Figure 9.** SOM evaluations of test samples. (a) ,(b) two different bruised apples, c) Non-bruised apple. Bruised area is colored as bright yellow